\documentclass[amsmath,amssymb,11pt]{revtex4-2}
\usepackage{appendix}
\usepackage{booktabs}
\usepackage{dcolumn}
\usepackage{float}
\usepackage{graphicx}
\usepackage[colorlinks,linkcolor=blue,anchorcolor=blue,urlcolor=blue,citecolor=blue]{hyperref}
\usepackage{makecell}
\usepackage{multirow}
\usepackage{subfigure}
\usepackage{xcolor}
\usepackage{xfrac}
\usepackage{amsmath}
\usepackage{orcidlink}
\begin{document}

\title{Energy Extraction from Rotating Charged Black Holes in Kalb-Ramond Gravity}

\vspace{0.25in}
\author{Jin-Tao Yao\orcidlink{0009-0001-2013-3578}$^1$}

\author{Ke-Jian He\orcidlink{0000-0002-1408-0019}$^2$}

\author{Zi-Chao Lin\orcidlink{0000-0002-7689-1749}$^1$}

\author{Hao Yu\orcidlink{0000-0003-1689-2421}$^1$}
	\email[Corresponding author: ]{yuhaocd@cqu.edu.cn}

\affiliation{$^1$Department of Physics and Chongqing Key Laboratory for Strongly Coupled Physics, Chongqing University, Chongqing
401331, China}
\affiliation{$^2$Department of Mechanics, Chongqing Jiaotong University, Chongqing 400074, China}

\vspace{0.25in}
\begin{abstract}

This work presents a comprehensive study of energy extraction via the Comisso-Asenjo magnetic reconnection mechanism from rotating charged black holes in the context of Kalb-Ramond (KR) gravity. We systematically investigate the influence of various parameters on the energy extraction process, comparing the results in two distinct regions: the circular orbit region and the plunging region. {The results reveal that the Lorentz-violating parameter has a significant impact on energy extraction, affecting not only the parameter space where energy extraction is possible, but also the energy extraction power and efficiency.}
It is found that the energy extraction process in the circular orbit region can offer a promising avenue for constraining KR gravity. In contrast, although energy extraction from the plunging region remains feasible even for black holes with relatively low spins and takes place nearer to the event horizon, its sensitivity to the Lorentz-violating parameter is significantly reduced. Overall, the Comisso-Asenjo magnetic reconnection mechanism can serve as a probe of the KR field, particularly through the energy extraction process in the circular orbit region.
\end{abstract}

\maketitle

\section{Introduction}
The quest to extract the rotational energy stored in astrophysical black holes is one of the most intriguing topics in theoretical physics and high-energy astrophysics. Since Roger Penrose's seminal proposal that energy could, in principle, be extracted from a rotating Kerr black hole via a special mechanism~\cite{Penrose1969pc}, this field has evolved from a theoretical curiosity into a cornerstone for understanding active galactic nuclei and relativistic jets~\cite{Blandford:1982xxl,Tchekhovskoy:2011zx,McKinney:2012vh,Ghisellini:2014pwa,Blandford:2018iot,Narayan:2011eb,Zamaninasab:2014dhz}. The Penrose process hinges crucially on the existence of an ergosphere: a region outside the event horizon where the spacetime frame-dragging is so severe that all trajectories are forced to corotate with the black hole. Within this ergosphere, a particle can split into two, with one fragment carrying negative energy (as measured from infinity) into the horizon, thereby reducing the black hole mass and angular momentum, while the other fragment escapes to infinity with more energy than the original particle~\cite{Penrose1971,misner1973gravitation}.

Subsequent decades have seen the proposal of various alternative and potentially more efficient mechanisms for tapping into the black hole's rotational energy. The Blandford-Znajek mechanism, which describes the electromagnetic extraction of energy via a magnetic field in an accretion disk, has become a mainstream model for powering relativistic jets~\cite{Blandford:1977ds}. While the Blandford-Znajek process extracts rotational energy directly from a spinning black hole, the Blandford-Payne mechanism~\cite{Blandford:1982xxl} operates in the accretion disk, driving magnetocentrifugal outflows that are believed to dominate jet production in systems with modest black hole spin or thick disks. More recently, the Banados-Silk-West mechanism demonstrated that particle collisions near the event horizon of an extremal Kerr black hole could result in arbitrarily high center-of-mass energies, effectively casting the black hole as a natural particle accelerator~\cite{Banados:2009pr}.
{One of the most recent developments is the Comisso-Asenjo magnetic reconnection mechanism, proposed in 2021~\cite{Comisso2021}.} This mechanism posits that within the highly magnetized plasma in the ergosphere, relativistic magnetic reconnection can spontaneously generate plasma outflows. Crucially, under strong frame-dragging, one set of plasmoids can acquire negative energy at infinity, while the other is violently accelerated outward, thus converting the black hole's rotational energy into kinetic energy of the escaping plasma. The magnetic reconnection mechanism offers a compelling and potentially highly efficient pathway for energy extraction.

In recent years, the exploration of energy extraction via the magnetic reconnection mechanism has developed far beyond the standard Kerr solution of general relativity (GR). These developments usually result in modifications to critical parameters of black holes, including the size of the ergosphere, the location of the innermost stable circular orbit (ISCO), and the photon sphere, which significantly impacts the energy extraction process~\cite{Khodadi:2022dff,Carleo:2022qlv,Wei2022,Chen2024,Ye:2023xyv,Zhang:2024ptp,Rodriguez:2024jzw,Shen:2024exn,Wang:2025pqh,Rueda:2023mtp,Khodadi:2023juk,Shaymatov:2023dtt,zhao2025,YuChih:2025hsg,Ortiqboev:2024mtk}. For instance, a negative cosmological constant extends the radial range of magnetically dominated reconnection where the energy extraction condition holds, thereby enabling energy extraction from black holes with lower spin~\cite{zhao2025}. In bumblebee gravity, when the Lorentz symmetry breaking parameter is negative, magnetic reconnection serves as a more efficient mechanism for energy extraction compared to the Blandford-Znajek mechanism~\cite{Khodadi:2022dff}, which has also been confirmed in similar studies~\cite{Carleo:2022qlv,YuChih:2025hsg,Ortiqboev:2024mtk}. Accounting for the backreaction of the magnetic field on the spacetime geometry, although stronger magnetic fields enhance plasma magnetization and facilitate energy extraction, this benefit is progressively offset by an increasingly inhibitory backreaction. Balancing these competing effects shows that a moderate magnetic field strength is optimal for energy extraction~\cite{Zhang:2024ptp}. Furthermore, studies on energy extraction from black holes via magnetic reconnection in different environments have also attracted widespread attention, such as in the presence of dark matter~\cite{Rodriguez:2024jzw} and extra dimensions~\cite{Wei2022}.

Within this rich landscape of modified gravity, Kalb-Ramond (KR) gravity stands out for its fundamental motivation. The KR field is an antisymmetric tensor field with two indices appearing in string theory, and a non-zero vacuum expectation value for this field spontaneously breaks Lorentz invariance, introducing a signature of quantum gravity at low energies~\cite{Kalb:1974yc,Losev:2005pu,Gaona:2006td,Gross:1984dd}. Despite the stringent constraints imposed by solar system tests~\cite{Kar:2002xa,Yang2023}, the KR field continues to be a focus of extensive research in numerous domains of gravitational research over the years. The duality of the KR and Proca fields is frequently found in the literature~\cite{Kawai:1980qq,Hell:2021wzm,Quevedo:1996uu,Quevedo:1997jb,DAuria:2004psr,Dalmazi:2011df,DeGracia:2017spm,Shifman:2018adu}. In extra-dimensional theories, the KR field has been investigated as a potential probe of extra dimensions~\cite{Chakraborty:2016lxo}, with attention also given to its localization properties~\cite{Du:2013bx,Cruz:2009ne} and other associated issues~\cite{DeRisi:2007dn,Chakraborty2018,Erdmenger:2007bn,Mukhopadhyaya:2004cc}. The possibility of inflation with the KR field coupled to gravity was investigated in Refs.~\cite{Elizalde:2018now,Aashish:2018lhv}. Similar applications of the KR field in cosmology also include a range of issues such as gravitational lensing~\cite{Kar:2002xa,Chakraborty:2016lxo,Ovgun:2025ctx,Mangut:2025gie,Junior:2024vdk,VP:2025lba}, dark matter~\cite{Capanelli:2023uwv,Dashko:2018dsw}, and leptogenesis~\cite{deCesare:2014dga}.

In the context of black hole research, the KR field has played an increasingly important role, with significant achievements in areas such as black hole shadow~\cite{Jha:2024xtr,Junior:2024ety,Muhammad2024,Zeng:2025kyv,Yang:2025whw,Yang:2025byw,Xu:2025iwg,Liu:2024lve}, geodesic motion of particles~\cite{AraujoFilho:2024rcr,jumaniyozov2025,Jumaniyozov:2025wcs,Jumaniyozov:2024eah}, and quasi-normal modes~\cite{Guo:2023nkd,Filho:2023ycx}. The degree of Lorentz violation in KR gravity is characterized by a dimensionless parameter (denoted as the Lorentz-violating parameter), which can directly modify the spacetime geometry of black holes and consequently influence their essential features (e.g., the radii of the event horizon, ergosphere, and ISCO)~\cite{Lessa2019,Yang2023,Bluhm2007,Liu2024,Liu2025,Muhammad2024,jumaniyozov2025,Duan:2023gng,Muhammad2024,jumaniyozov2025,Kumar:2020hgm}. However, the extent to which the effect of the Lorentz-violating parameter on energy extraction can be observed remains a question with a lack of research. This work aims to fill this gap by presenting a comprehensive analysis of energy extraction via the Comisso-Asenjo magnetic reconnection mechanism from a rotating charged black hole in KR gravity. We conduct research across two critical regions of charged rotating black holes: the circular orbit region and the plunging region~\cite{Chen2024,Shen:2024sdr,Cheng:2025qlc}. {We present a systematic analysis of the effects  of  relevant parameters on energy extraction, with a focus on the Lorentz-violating parameter related to the KR field.}

The paper is organized as follows. In section~\ref{sec.2}, we simply introduce the rotating charged black hole in KR gravity. Section~\ref{sec.3} is dedicated to the analysis of energy extraction via the Comisso-Asenjo magnetic reconnection mechanism in the circular orbit region, detailing the parameter space for the allowed energy extraction region, the energy extraction power and efficiency. We mainly focus on the influence of the Lorentz-violating parameter on energy extraction. Section~\ref{sec.4} extends the similar analysis to the plunging region. Finally, we present our conclusion in Section~\ref{sec.5}.

\section{Rotating Charged Black Holes in Kalb-Ramond Gravity}\label{sec.2}

In this section, we review the exact solution and geodesic equations for rotating charged black holes in KR gravity. We consider the Einstein-Hilbert action non-minimally coupled to the KR field, incorporating self-interaction potential breaking Lorentz and diffeomorphism symmetry~\cite{Altschul2010}:
\begin{align}
S =& \frac{1}{16\pi G} \int d^4 x \sqrt{-g} \Bigg[ R - 2\Lambda - \frac{1}{6} H^{\mu\nu\rho} H_{\mu\nu\rho} - V(B^{\mu\nu} B_{\mu\nu}\pm b^{\mu\nu} b_{\mu\nu}) \nonumber \\
&+ \xi_2 B^{\rho\mu} B^{\nu}{}_{\mu} R_{\rho\nu} + \xi_3 B^{\mu\nu} B_{\mu\nu} R \Bigg] + S_{\text{M}}.
\end{align}
Here, $\Lambda$ is the cosmological constant,
$H_{\mu\nu\rho} = \partial_{[\mu} B_{\nu\rho]}$ is the field strength of the KR field, and the parameters
$\xi_2$ and $\xi_3$ are independent non-minimal coupling constants between the KR field and curvature.
$V(B^{\mu\nu} B_{\mu\nu}\pm b^{\mu\nu} b_{\mu\nu})$ is the self-interaction potential depending on the vacuum configuration of the KR field. It is worth noting that variations in the vacuum configuration give rise to correspondingly different black hole solutions~\cite{Lessa2019,Yang2023,Bluhm2007,Liu2024,Liu2025,Muhammad2024,jumaniyozov2025,Duan:2023gng,Muhammad2024,Kumar:2020hgm}. In Ref.~\cite{Lessa2019}, L. A. Lessa {\it et al}. considered the KR field in the vacuum configuration $B^{\mu\nu} B_{\mu\nu} = b^{\mu\nu} b_{\mu\nu}$, and obtained an exact parameter dependent power-law black hole solution, which is static, neutral, and spherically symmetric. This solution can mimic a Reissner-Nordstr$\ddot{\text{o}}$m black hole under specific parameter choices, despite the absence of electric charge. However, in Ref.~\cite{Yang2023}, under the same black hole configuration and gravitational setup, K. Yang {\it et al}. adopted a different vacuum configuration (i.e., $V' = 0$ indicating that the vacuum expectation value is located at the local minimum of the potential~\cite{Bluhm2007}), which led to a formally distinct solution. This solution was later extended to more general cases involving static and neutral black holes~\cite{Liu2024,Liu2025}. In this work, we employ the vacuum configuration studied by K. Yang {\it et al}.~\cite{Yang2023}, assuming a vanishing cosmological constant. With this premise, the rotating charged black hole in Boyer-Lindquist coordinates with geometric units can be described by the following line element~\cite{Muhammad2024,jumaniyozov2025}:
\begin{align}\label{metric}
	ds^2 = &-\frac{\Delta - a^2 \sin^2\theta}{\Sigma} dt^2 + \frac{\Sigma}{\Delta} dr^2 + \Sigma d\theta^2	\nonumber	\\
	&+ \frac{\sin^2\theta}{\Sigma} \left[ \left( r^2 + a^2 \right)^2 - a^2 \Delta \sin^2\theta  \right] d\varphi^2	\nonumber	\\
	&- \frac{2 a \sin^2\theta }{\Sigma} \left( r^2 + a^2 - \Delta \right) dt d\varphi ,
\end{align}
where the special symbols are defined as
\begin{align}
	\Delta &= \frac{r^2}{1 - \ell} - 2 M r + a^2 + \frac{Q^2}{\left( 1 - \ell \right)^2} ,	\\
	\Sigma &= r^2 + a^2 \cos^2\theta .
\end{align}
The parameters $M$, $a$, and $Q$ denote the black hole mass, spin, and (electric) charge, respectively. The dimensionless parameter
$\ell$ characterizes the strength of Lorentz violation caused by the non-zero vacuum expectation value of the KR field. When $\ell = 0$, this black hole solution reduces to the standard Kerr-Newman solution.

The horizons of the black hole are located at surfaces where the radial component of the metric diverges. For the metric~\eqref{metric}, this divergence condition  corresponds to $\Delta = 0$, which yields
\begin{equation}\label{horizons}
	r_\pm = M(1-\ell) \pm \sqrt{M^2 (1-\ell)^2 - a^2 (1 - \ell) - \frac{Q^2}{1-\ell}}.
\end{equation}
In general, one may define the event horizon as $r_{\rm H} = r_+$ and the Cauchy horizon as $r_C = r_-$, provided that the equation $\Delta = 0$ admits two real roots. On the other hand, the infinite redshift surfaces are given by the condition $g_{tt}=0$, from which, one can obtain
\begin{equation}\label{ergosphere}
	r_\pm^s = M(1-\ell) \pm \sqrt{M^2 (1-\ell)^2 - a^2 \cos^2\theta (1 - \ell) - \frac{Q^2}{1-\ell}} .
\end{equation}
In the following, we investigate energy extraction from the black hole via the Comisso-Asenjo magnetic reconnection mechanism, thus focusing on the outer infinite redshift surface, denoted by $r_{\rm E} = r_+^s$. The space region bounded by the event horizon $r_{\rm H}$ and the outer infinite redshift surface $r_{\rm E}$ is known as the ergosphere, where energy extraction becomes possible. Following the work of L. Comisso {\it et al}.~\cite{Comisso2021}, we model the magnetic reconnection process as occurring within a current sheet located on the equatorial plane. Under the assumption, the outer infinite redshift surface reduces to
\begin{equation}
	r_{\rm E} = M(1-\ell) + \sqrt{M^2 (1-\ell)^2 - \frac{Q^2}{1-\ell}}.
\end{equation}

In a stationary, axially symmetric spacetime, the norm of the four-velocity of a test particle is a constant ($-m^2$) along geodesics due to parallel transport. Moreover, the energy $E$ and $z$-component angular momentum $L$ about the symmetry axis are also conserved along geodesic. For the model that the current sheet only moves on the equatorial plane of the black hole, these three constants are sufficient to describe the geodesic equations of the test particle around the black hole. Following Ref.~\cite{Shaymatov:2023dtt} and employing the Hamilton-Jacobi equation~\cite{Carter1968}, the first-order geodesic equations can be written as
\begin{align}
	\Sigma \dot{t} =\,& \frac{(r^2 + a^2) \left[ (r^2 + a^2) E - a L \right]}{\Delta} - a \left( a E - L \right), \label{dt}	\\
	\Sigma^2 \dot{r}^2 \equiv\,& R(r)=\left[ \left( r^2 + a^2 \right)E - a L \right]^2 - \Delta \left[ (aE-L)^2  + m^2 r^2 \right], \label{dr}	\\
	\Sigma \dot{\varphi} =\,& \frac{a \left[ (r^2 + a^2) E - a L \right]}{\Delta} - \left( a E - L \right) ,	\label{dvarphi}
\end{align}
where the dot denotes the derivative with respect to the affine parameter. Note that in this work $m^2 = 1$ for massive particles and $m^2 =0$ for massless ones, respectively. With the definition of $R(r)$, the radius of the photon sphere can be calculated by
\begin{equation}
	R(r_{\rm PH})=0,\qquad	R'(r_{\rm PH})=0,\qquad \text{with}\,\,\, m^2=0,
\end{equation}
where the prime denotes the derivative with respect to $r$. The radius of the ISCO can be obtained by solving
\begin{equation}
	R(r_{\rm ISCO})=0 ,	\qquad	R'(r_{\rm ISCO})=0 ,	\qquad	R''(r_{\rm ISCO})=0, \qquad\text{with}\,\,\, m^2=1.
\end{equation}
For retrograde orbits, the photon sphere lies outside the ergosphere, making energy extraction impossible under the circular orbit region. Regarding the plunging region, although particles can enter the ergosphere from the ISCO, the energy density per enthalpy at infinity for both accelerated/decelerated plasma remains greater than zero, which also prevents energy extraction~\cite{Chen2024}. Therefore, in practical applications, energy extraction mechanisms primarily focus on prograde orbits. For test particles on prograde equatorial orbits, the corresponding Keplerian angular velocity, as measured by an observer at infinity, is given by
\begin{equation}
	\Omega_K \equiv \frac{d\,\varphi}{d\,t} = \frac{\sqrt{M \,r (1-\ell)^2 - Q^2}}{a\sqrt{M\, r (1-\ell)^2 - Q^2} + r^2 (1-\ell)}.
\end{equation}
As anticipated, the Keplerian angular velocity is determined entirely by the azimuthal angular velocity, with no contribution from the radial component.

\section{Energy Extraction from the Circular Orbit Region}\label{sec.3}
Building upon the preceding groundwork, we now investigate energy extraction from the circular orbit region around rotating charged black holes via the Comisso-Asenjo magnetic reconnection mechanism  within the framework of KR gravity, and analyze how various parameters influence the energy extraction process. We primarily focus on the Lorentz-violating parameter $\ell$, which may serve as a key diagnostic for determining whether the KR field is present in the spacetime. To facilitate the analysis of the plasma mass-energy density in magnetic reconnection, it is common to introduce a locally non-rotating frame known as the zero-angular-momentum observer (ZAMO) frame~\cite{Bardeen1972}. The advantage of employing the ZAMO frame lies in its ability to render the spacetime locally Minkowskian from the observer's perspective, thereby simplifying our calculations and analyses. In particular, within the ZAMO frame, the line element takes the form
\begin{equation}
	ds^2 = -\alpha^2 dt^2 + \sum_{i=1}^{3}(\sqrt{g_{ii}} dx^i - \alpha \,\beta^i dt)^2	,
\end{equation}
where $\alpha$ is the lapse function and $\beta^i = \sqrt{g_{ii}} \omega^i / \alpha$ is the shift vector with $\omega^i$ being the velocity of frame-dragging. They take the form
\begin{equation}
	\alpha = \sqrt{\frac{g_{t\varphi}^2}{g_{\varphi\varphi}} - g_{tt}} ,	\qquad	\beta^i = \left( 0, 0, \beta^\varphi \right) ,	\qquad	\omega^i = \left( 0, 0, -\frac{g_{t\varphi}}{g_{\varphi\varphi}} \right) .
\end{equation}
Thus, the corotating Keplerian velocity in the ZAMO frame is given by~\cite{Wei2022,Wang2022}
\begin{equation}
	\hat{v}_K = \frac{\sqrt{g_{\varphi\varphi}}}{\alpha} \left( \Omega_K - \omega^\varphi \right).
\end{equation}

Following Ref.~\cite{Comisso2021}, the plasma is modeled using the one-fluid approximation, wherein it is characterized by specific particle density and pressure. In this case, the energy-momentum tensor  of the plasma can be expressed as
\begin{equation}
	T^{\mu\nu} = p \,g^{\mu\nu} + w\, u^\mu u^\nu + F^{\mu}_{~\delta} F^{\nu \delta} - \frac{1}{4} g^{\mu \nu} F^{\alpha \beta} F_{\alpha \beta} ,
\end{equation}
where $p$, $w$, $u^\mu$, and $F^{\mu\nu}$ are the proper plasma pressure, enthalpy density, four-velocity, and electromagnetic field tensor, respectively. We assume that magnetic reconnection is highly efficient in converting magnetic energy into kinetic energy, such that the electromagnetic energy at infinity becomes negligible, and the hydrodynamic energy dominates. Owing to the strong gravitational field near the black hole, relativistic effects must typically be taken into account when analyzing the behavior of the plasma. For a relativistic hot plasma, the enthalpy density and proper pressure satisfy $w = 4p$~\cite{Zeng2025a}. Under the assumptions that the plasma is incompressible and adiabatic~\cite{Koide2008}, and that the thermal pressure gradient along the outflow direction is negligible compared to the magnetic tension force~\cite{Liu2017}, the hydrodynamic energy per unit enthalpy of the plasma at infinity can be expressed as~\cite{Comisso2021}
\begin{align}\label{CO_plasma_energy}
	e_{\pm}^\infty = -\frac{\alpha g_{\mu 0} T^{\mu 0}}{w}
	= \alpha \hat{\gamma}_K \gamma_{\rm out} \left[1 + \beta^\varphi \hat{v}_K \pm v_{\rm out} \left( \hat{v}_K + \beta^\varphi\right) \cos\xi  \right] - \frac{\alpha}{4 \hat{\gamma}_K \gamma_{\rm out} \left( 1 \pm \hat{v}_K v_{\rm out} \cos\xi \right)}.
\end{align}
Here, $\hat{\gamma}_K = \frac{1}{\sqrt{1 - \hat{v}_K^2}}$ is the Lorentz factor corresponding to the corotating Keplerian velocity $\hat{v}_K$ in the ZAMO frame, and $\xi$ is the orientation angle between the magnetic field lines and the azimuthal direction in the equatorial plane. In the local rest frame of the bulk plasma, the outflow velocity can be written as~\cite{Comisso2021}
\begin{equation}
	v_{\rm out} = \sqrt{\frac{ \sigma }{ 1 + \sigma }},
\end{equation}
where $\sigma$ is the plasma magnetization upstream of the reconnection layer. Therefore, $\gamma_{\rm out} = \sqrt{ 1 + \sigma}$ is the Lorentz factor corresponding to $v_{\rm out}$. The ``$+$'' and ``$-$'' signs in Eq.~\eqref{CO_plasma_energy} represent outflows in the corotating and counterrotating directions, respectively, as observed in the local rest frame of the bulk plasma.

According to the original Penrose process~\cite{Penrose1971}, energy extraction from a black hole is feasible under two conditions: (1) decelerated particles within the ergosphere acquire negative energy, and (2) accelerated particles escape to infinity. In the context of energy extraction via the magnetic reconnection mechanism, the hydrodynamic energy per unit enthalpy must satisfy the following two conditions:
\begin{equation}\label{conditions_of_extract_energy}
	e_-^\infty < 0 ,	\qquad	\Delta e_+^\infty = e_+^\infty - \left( 1 - \frac{\Gamma}{\Gamma - 1} \frac{p}{w} \right) = e_+^\infty > 0,
\end{equation}
where $\Gamma= 4/3$ is the polytropic index for a relativistic hot plasma~\cite{Comisso2021}. With the conditions and the exact black hole solution~\eqref{metric}, we can proceed to analyze the dependence of energy extraction on the model parameters in KR gravity.

\subsection{Parameter Space for Energy Extraction}
In the present model, the hydrodynamic energy per unit enthalpy of the plasma at infinity is determined by several parameters: the black hole mass $M$, spin $a$, charge $Q$, the radial location $r$ of the dominant magnetic reconnection point (denoted as X-point), the plasma magnetization $\sigma$, the orientation angle $\xi$, and the Lorentz-violating parameter $\ell$. The effects of these parameters on energy extraction are analyzed in detail below. In this work, we focus on the (allowed) energy extraction region within the parameter space spanned by $\{a/M, r/M\}$. For clarity, when analyzing the impact of a single parameter on energy extraction, all other parameters are held constant at representative values.

\begin{figure*}[t]
	\centering
	\subfigure[]{\includegraphics[width=0.45\linewidth]{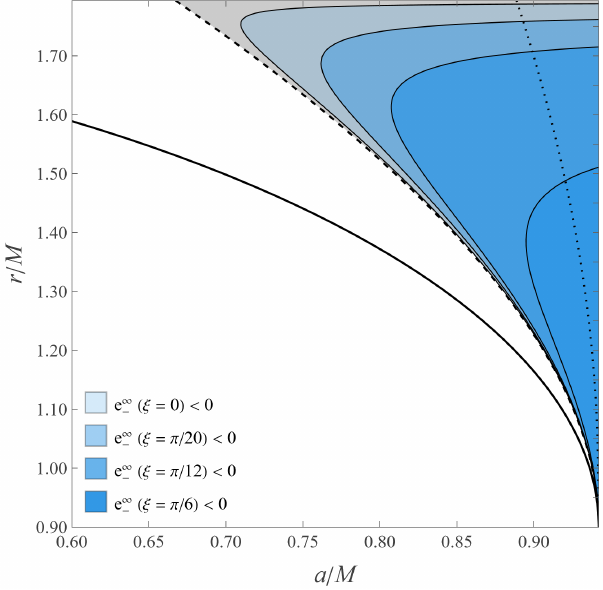}\label{CO_Region_different_xi}}
	\subfigure[]{\includegraphics[width=0.45\linewidth]{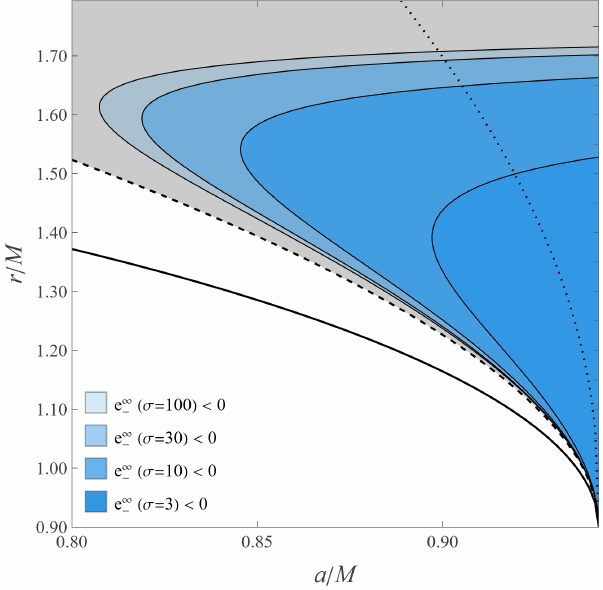}\label{CO_Region_different_sigma}}
	\\
	\subfigure[]{\includegraphics[width=0.45\linewidth]{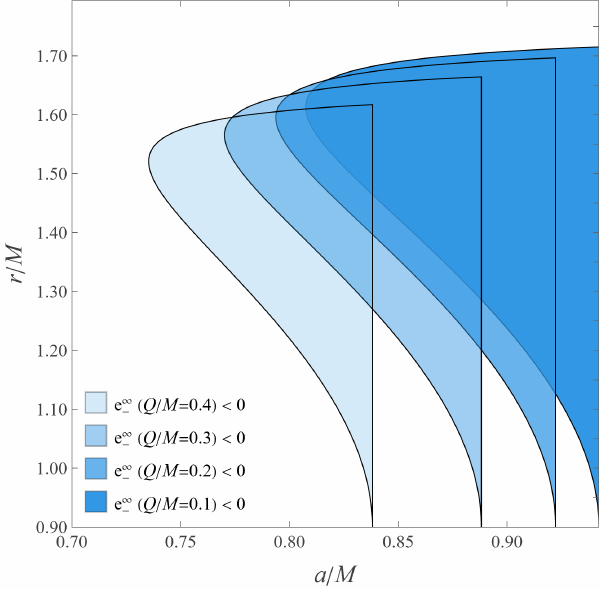}\label{CO_Region_different_Q}}
	\subfigure[]{\includegraphics[width=0.45\linewidth]{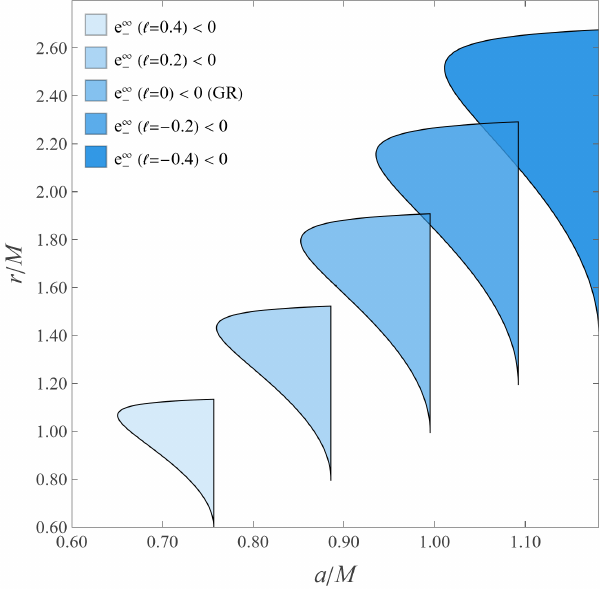}\label{CO_Region_different_ell}}
	\caption{Energy extraction region in the circular orbit region for different parameter combinations. In each panel, the fixed and variable parameters are set to the following values: (a) $\ell=0.1$, $Q/M=0.1$, $\sigma=100$, $\xi = \{ 0, \pi/20, \pi/12, \pi/6 \}$; (b) $\ell=0.1$, $Q/M=0.1$, $\xi=\pi/12$, $\sigma = \{ 3, 10, 30, 100 \}$; (c) $\ell=0.1$, $\sigma=100$, $\xi=\pi/12$, $Q/M = \{ 0.1, 0.2 , 0.3, 0.4\}$; (d) $Q=0.1$, $\sigma=100$, $\xi=\pi/12$, $\ell = \{ -0.4, -0.2, 0, 0.2, 0.4 \}$. }
	\label{CO_Region}
\end{figure*}

In Fig.~\ref{CO_Region}, we show the  energy extraction region for different parameter combinations, including the orientation angle $\xi$, the plasma magnetization $\sigma$, the black hole charge $Q$, and the Lorentz-violating parameter $\ell$. In Figs.~\ref{CO_Region_different_xi} and~\ref{CO_Region_different_sigma}, the solid, dashed, and dotted black lines represent the event horizon $r_{\rm H}$, the photon sphere radius $r_{\rm PH}$, and the ISCO radius $r_{\rm ISCO}$, respectively. In Figs.~\ref{CO_Region_different_Q} and~\ref{CO_Region_different_ell}, we have omitted the lines marking these characteristic radii, as their positions depend on both $Q$ and $\ell$, and plotting them for all parameter values would lead to visual overcrowding.

First, we study the effects of $\xi$ [see Fig.~\ref{CO_Region_different_xi}] and $\sigma$ [see Fig.~\ref{CO_Region_different_sigma}] on energy extraction. The gray area represents $\Delta e^\infty_+ >0$, and the blue area represents $e_-^\infty < 0$. Since the gray area always encompasses the blue area for all parameter values, the blue area directly represents the energy extraction region. It is observed that the energy extraction region contracts as $\xi$ increases, and expands with increasing $\sigma$. Therefore, for small values of $\xi$ (or large values of $\sigma$), the energy extraction region fully encompasses that corresponding to large $\xi$ (or small $\sigma$). These phenomena are consistent with findings reported in previous similar studies~\cite{Comisso2021,Wei2022,Wang2022,Zhang:2024rvk,Li:2023htz,Zeng2025b,Li:2023nmy,Shen:2024sdr,Long:2024tws}, as the influence of these two parameters on energy extraction is largely independent of the gravitational theory.

Next, we are concerned with the effects of $Q$ [see Fig.~\ref{CO_Region_different_Q}] and $\ell$ [see Fig.~\ref{CO_Region_different_ell}] on energy extraction. For any given parameter set, the gray area corresponding to $\Delta e^\infty_+ > 0$ continues to encompass the blue area where $e_-^\infty < 0$. Thus, for enhanced clarity and intuitiveness, we display only the blue area, which delineates the parameter space where energy extraction is possible. Because the charge $Q$ directly influences the black hole solution, its effect on energy extraction follows a qualitatively distinct trend compared to those of the parameters $\xi$ and $\sigma$. As illustrated in Fig.~\ref{CO_Region_different_Q}, the energy extraction region shrinks with increasing $Q/M$ and shifts toward smaller values of both $r/M$ and $a/M$. Therefore, for different charges, the parameter space for energy extraction does not fully overlap. Typically, these properties can also be observed in other charged black holes~\cite{Zeng2025a,YuChih:2025hsg,Shaymatov:2023dtt}. The influence of the KR field on energy extraction from rotating charged black holes is reflected in the Lorentz-violating parameter $\ell$. From Fig.~\ref{CO_Region_different_ell}, one can find that, similar to Fig.~\ref{CO_Region_different_Q}, the energy extraction region also shrinks as the Lorentz-violating parameter $\ell$ increases, and shifts toward smaller values of both $r/M$ and $a/M$. However, the influence of $\ell$ on energy extraction is more significant than that of the charge $Q$. When $\ell$ changes significantly (e.g., from $\ell = 0.2$ to $\ell = 0.4$, with $Q=0.1$, $\sigma=100$, and $\xi=\pi/12$ held fixed), the corresponding energy extraction regions are entirely non-overlapping. Moreover, in the case of $\ell = 0$ (i.e., GR), energy extraction is feasible only when the black hole spin $a/M$ lies approximately between 0.87 (lower bound) and 1 (upper bound). However, as $\ell$ increases (e.g., $\ell=0.4$), energy extraction becomes feasible even if $a/M$ is less than 0.87. Conversely, a larger spin becomes necessary for energy extraction as $\ell$ decreases. Hence, for a black hole with known mass, spin, and charge, the possibility of energy extraction could serve as a constraint on the Lorentz-violating parameter $\ell$, thereby providing a means to probe the KR field.

\subsection{Power and Efficiency}
We now evaluate the rate of energy extraction from the circular orbit region around rotating charged black holes in KR gravity. The energy extraction power $P_{\rm extra}$ carried by the escaping plasma from a black hole is defined as~\cite{Comisso2021}
\begin{equation}\label{power}
	P_{\rm extra} = - w \,e^\infty_- A_{\rm in} U_{\rm in},
\end{equation}
where $U_{\rm in}=\mathcal{O}(10^{-1})$ in the collisionless regime and $U_{\rm in} = \mathcal{O}(10^{-2})$ in the collisional regime. We adopt the collisionless regime in this work. $A_{\rm in}$ is the cross sectional area of the inflowing plasma, which, for the sake of simplicity, can be expressed as~\cite{Comisso2021}
\begin{equation}\label{cross_sectional_area}
	A_{\rm in} \sim r_{\rm E}^2 - r_{\rm PH}^2.
\end{equation}
Recall that $r_{\rm E}$ and $r_{\rm PH}$ represent the ergosphere boundary and the photon sphere radius, respectively.

\begin{figure*}[t]
	\centering
	\subfigure[]{\includegraphics[width=0.45\linewidth]{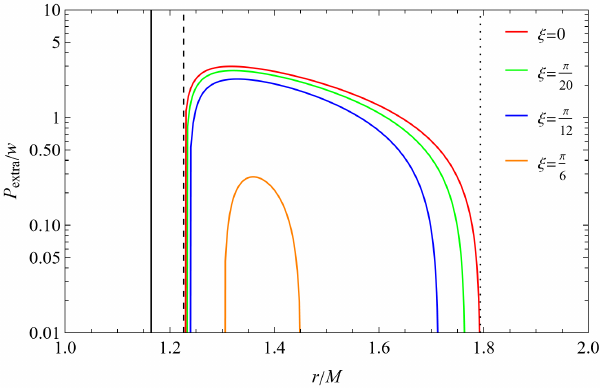}\label{CO_Power_different_xi}}
	\subfigure[]{\includegraphics[width=0.45\linewidth]{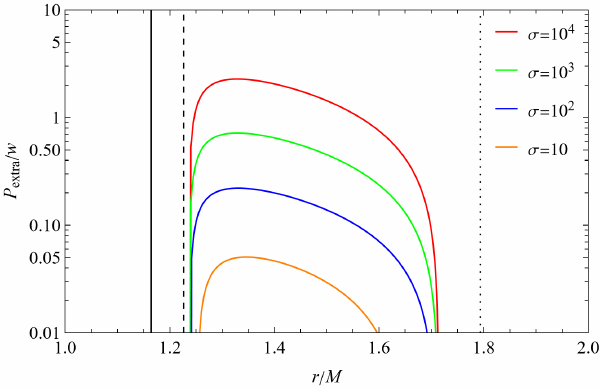}\label{CO_Power_different_sigma}}
	\\
	\subfigure[]{\includegraphics[width=0.45\linewidth]{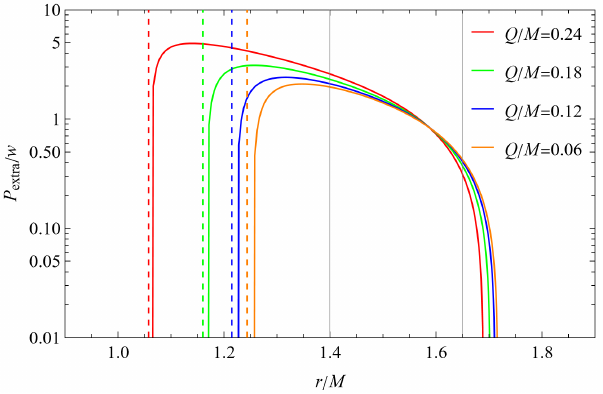}\label{CO_Power_different_Q}}
	\subfigure[]{\includegraphics[width=0.45\linewidth]{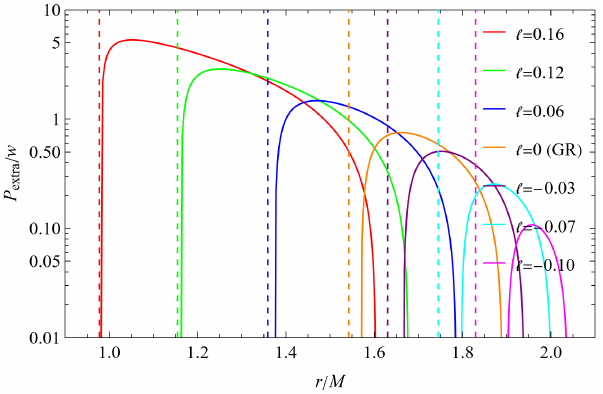}\label{CO_Power_different_ell}}
	\caption{Energy extraction power per unit enthalpy density $P_{\rm extra}/w$ in the circular orbit region for different parameter combinations. The vertical dashed lines mark the position of the photon sphere. In each panel, the black hole spin is fixed as $a/M=0.9$ and the other parameters are set to the following values: (a) $\ell=0.1$, $Q/M=0.1$, $\sigma=10^4$, $\xi = \{ 0, \pi/20, \pi/12, \pi/6 \}$; (b) $\ell=0.1$, $Q/M=0.1$, $\xi=\pi/12$, $\sigma = \{ 10, 10^2, 10^3, 10^4 \}$; (c) $\ell=0.1$, $\sigma=10^4$, $\xi=\pi/12$, $Q/M = \{ 0.06, 0.12 , 0.18, 0.24 \}$; (d) $Q/M=0.1$, $\sigma=10^4$, $\xi=\pi/12$, $\ell = \{-0.10, -0.07, -0.03, 0, 0.06, 0.12 , 0.16 \}$.}
	\label{CO_Power}
\end{figure*}

In Fig.~\ref{CO_Power}, we present the energy extraction power per unit enthalpy density $P_{\rm extra}/w$ (hereafter referred to as power) as a function of the radial location of the X-point (i.e., $r/M$ in the local rest frame of the bulk plasma) for different parameter combinations. The power exhibits a consistent trend across all parameter sets: it increases to a maximum before decreasing. In Figs.~\ref{CO_Power_different_xi} and~\ref{CO_Power_different_sigma}, the solid, dashed, and dotted black lines represent the locations of $r_{\rm H}$, $r_{\rm PH}$, and $r_{\rm E}$, respectively. One can find that the power increases monotonically with decreasing $\xi$ and increasing $\sigma$ for a fixed X-point. Moreover, as the parameter $\xi$ decreases (or equivalently, as $\sigma$ increases), the X-point corresponding to the maximum power shifts progressively closer to the location of the photon sphere. The influence of these two parameters on the power is completely monotonic, similar to their effects on the energy extraction region. Figs.~\ref{CO_Power_different_Q} and~\ref{CO_Power_different_ell} present how the power varies with the parameters $Q$ and $\ell$, respectively. The colored dashed lines indicate the location of the photon sphere radius for each set of parameters. For clarity, the locations of the event horizon and the ergosphere boundary are not shown in the figures. The influence of these two parameters on the power differs significantly from that of the parameters $\xi$ and $\sigma$. From Fig.~\ref{CO_Power_different_Q}, one can find that when the radial location of the X-point is relatively large (e.g., $r/M = 1.65$; refer to the gray solid line on the right), the power decreases with increasing $Q$. But, when the radial location of the X-point is small (e.g., $r/M = 1.4$; refer to the gray solid line on the left), the power increases as $Q$ increases. This indicates that the effect of $Q$ on the power is non-monotonic and depends on the X-point. Similar behavior is observed for the Lorentz-violating parameter $\ell$, but with more pronounced variations. For different values of $\ell$, both the maximum value of the power and the effective range of $r/M$ may differ significantly. For instance, comparing $\ell = 0$ and $\ell = -0.1$, the effective ranges of $r/M$ are entirely disjoint. This strong sensitivity of the power to the Lorentz-violating parameter $\ell$ may provide a promising avenue to constrain the KR field from the perspective of energy extraction from black holes.

\begin{figure*}[t]
	\centering
	\subfigure[]{\includegraphics[width=0.45\linewidth]{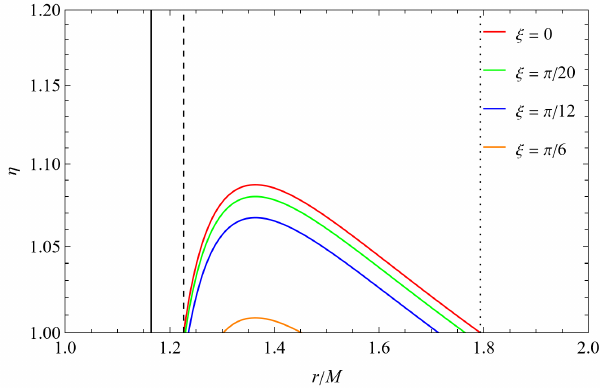}\label{CO_Eff_different_xi}}
	\subfigure[]{\includegraphics[width=0.45\linewidth]{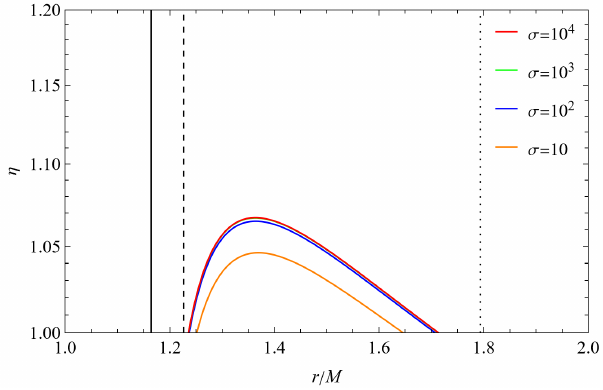}\label{CO_Eff_different_sigma}}
	\\
	\subfigure[]{\includegraphics[width=0.45\linewidth]{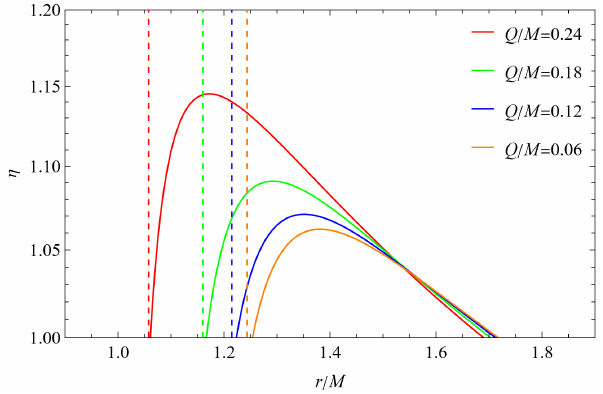}\label{CO_Eff_different_Q}}
	\subfigure[]{\includegraphics[width=0.45\linewidth]{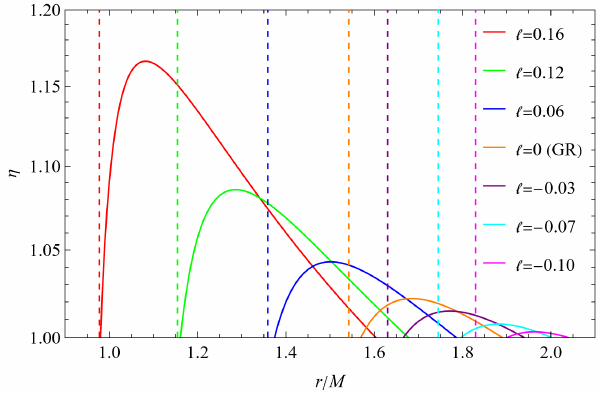}\label{CO_Eff_different_ell}}
	\caption{Energy extraction efficiency $\eta$ in the circular orbit region for different parameter combinations. The vertical dashed lines mark the position of the photon sphere. In each panel, the black hole spin is fixed as $a/M=0.9$ and the other parameters are set to the following values: (a) $\ell=0.1$, $Q/M=0.1$, $\sigma=10^4$, $\xi = \{ 0, \pi/20, \pi/12, \pi/6 \}$; (b) $\ell=0.1$, $Q/M=0.1$, $\xi=\pi/12$, $\sigma = \{ 10, 10^2, 10^3, 10^4 \}$; (c) $\ell=0.1$, $\sigma=10^4$, $\xi=\pi/12$, $Q/M = \{ 0.06, 0.12 , 0.18, 0.24 \}$; (d) $Q/M=0.1$, $\sigma=10^4$, $\xi=\pi/12$, $\ell = \{-0.10, -0.07, -0.03, 0, 0.06, 0.12 , 0.16 \}$.}
	\label{CO_Eff}
\end{figure*}

In the energy extraction process driven by magnetic reconnection, high-energy plasma outflows are generated by tapping into the rotational energy of the black hole. This mechanism requires an input of magnetic energy to redistribute the angular momentum of particles, enabling some to acquire negative energy at infinity while others are propelled outward. To quantify the effectiveness of this process, one can define the efficiency of extraction energy as
\begin{equation}\label{efficiency}
    \eta = \frac{e^\infty_+}{e^\infty_+ + e^\infty_-}.
\end{equation}
The condition $e^\infty_- < 0$ is a prerequisite for energy extraction. Thus, the resulting efficiency necessarily exceeds unity. Following the parameter combinations in Fig.~\ref{CO_Power}, we plot the energy extraction efficiency $\eta$ against the radial location of the X-point in Fig.~\ref{CO_Eff}. The overall trend is that the efficiency peaks then declines, which holds for all parameter combinations. A comparison of Figs.~\ref{CO_Power} and~\ref{CO_Eff} shows that all parameters influence the efficiency and the power in a similar manner. Hence, a repeated discussion of their effects on the efficiency is omitted here.

\section{Energy Extraction from the Plunging Region}\label{sec.4}

In this section, we investigate energy extraction via magnetic reconnection occurring in the plunging region~\cite{Chen2024,Shen:2024sdr,Cheng:2025qlc}. The bulk plasma initially follows circular orbits located outside the ISCO. Upon reaching the ISCO, it begins to plunge inward. Since orbits within the ISCO are dynamically unstable, any perturbation can induce a radial velocity, triggering the inward plunge. In the plunging region, the plasma acquires a non-negligible radial velocity component. Therefore, it is necessary to incorporate the relevant correction to Eq.~\eqref{CO_plasma_energy} to account for this effect. In this case, within the ZAMO frame, the four-velocity of the plasma is given by
\begin{equation}
	\hat{u}^\mu = \hat{\gamma}_s \{ 1, \hat{v}_s^{(r)}, 0, \hat{v}_s^{(\varphi)} \} = \left\{ \frac{E - \omega^\varphi L}{\alpha}, u^r \sqrt{g_{rr}}, 0, \frac{L}{\sqrt{g_{\varphi \varphi}}} \right\}	,
\end{equation}
where $u^r \equiv\, dr/d\tau$ is given by Eq.~\eqref{dr}. As the plasma plunges inward from the ISCO, its energy and angular momentum remain conserved. The hydrodynamic energy per unit enthalpy of the (corotating and counterrotating) plasma at infinity is given by~\cite{Chen2024}
\begin{equation}\label{PO_plasma_energy}
	e_{\pm}^\infty = \alpha \hat{\gamma}_s \gamma_{\rm out} \left[ 1 + \beta^\varphi \hat{v}_s^{(\varphi)} \pm v_{\rm out} \left( \hat{v}_s + \beta^\varphi \frac{\hat{v}_s^{(\varphi)}}{\hat{v}_s} \right) \cos\xi  \mp v_{\rm out} \beta^\varphi \frac{\hat{v}_s^{(r)}}{\hat{\gamma}_s \hat{v}_s} \sin\xi \right] - \frac{\alpha}{4 \hat{\gamma}_s \gamma_{\rm out} \left( 1 \pm \hat{v}_s v_{\rm out} \cos\xi \right)} ,
\end{equation}
where $\hat{v}_s = \sqrt{\left( \hat{v}_s^{(r)} \right)^2 + \left( \hat{v}_s^{(\varphi)} \right)^2}$ and $\hat{\gamma}_s$ is the Lorentz factor of $\hat{v}_s$. The criterion for energy extraction remains governed by Eq.~\eqref{conditions_of_extract_energy}, or equivalently, by the condition that the efficiency defined in Eq.~\eqref{efficiency} exceeds unity.

\begin{figure*}[t]
	\centering
	\subfigure[]{\includegraphics[width=0.45\linewidth]{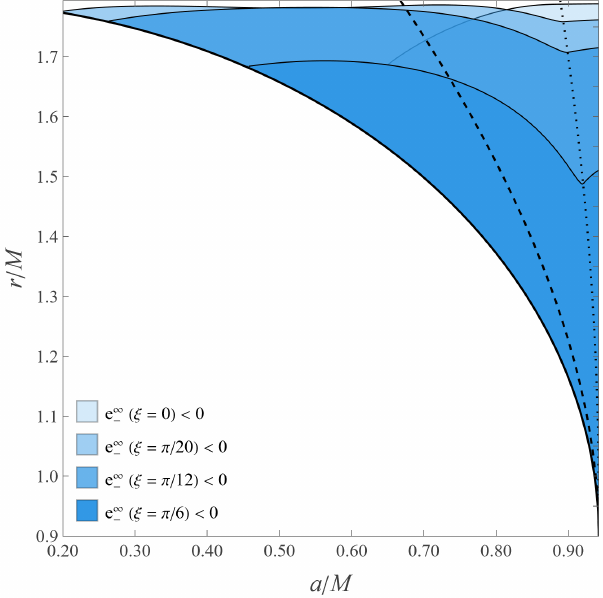}\label{PO_Region_different_xi}}
	\subfigure[]{\includegraphics[width=0.45\linewidth]{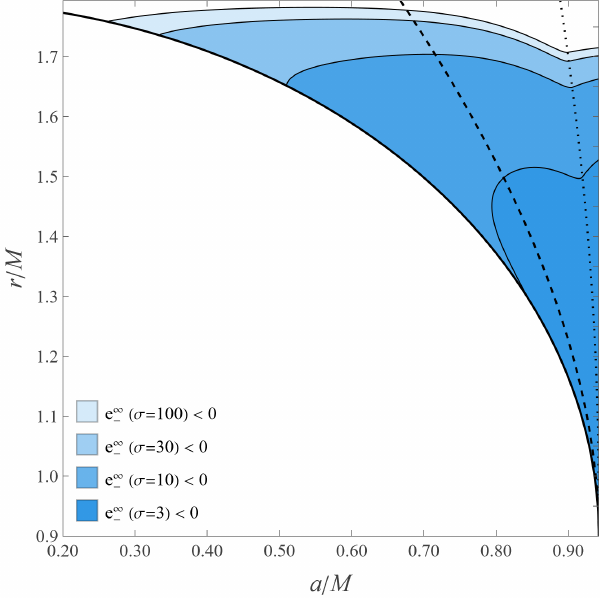}\label{PO_Region_different_sigma}}
	\\
	\subfigure[]{\includegraphics[width=0.45\linewidth]{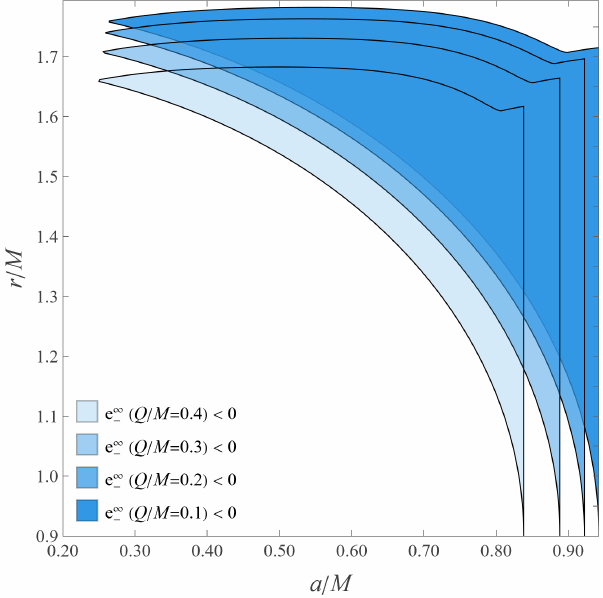}\label{PO_Region_different_Q}}
	\subfigure[]{\includegraphics[width=0.45\linewidth]{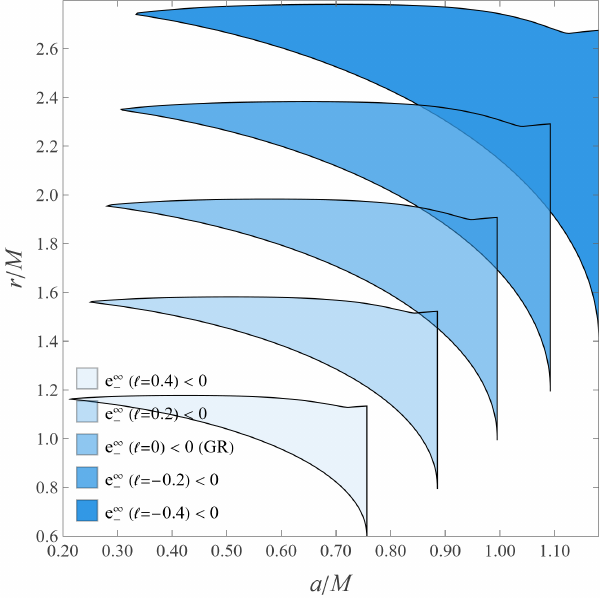}\label{PO_Region_different_ell}}
	\caption{Energy extraction region in the plunging region for different parameter combinations. The parameter combinations are identical to those used in Fig.~\ref{CO_Region}.}
	\label{PO_Region}
\end{figure*}

Following the steps outlined in the previous section, we first plot the energy extraction region for the plunging region in the parameter space $\{ a/M, r/M \}$. Adopting the same parameter combinations used in Fig.~\ref{CO_Region}, we present the results in Fig.~\ref{PO_Region}, where the blue area still represents the allowed energy extraction region. In Figs.~\ref{PO_Region_different_xi} and~\ref{PO_Region_different_sigma}, the solid, dashed, and dotted black lines still represent the event horizon $r_{\rm H}$, the photon sphere radius $r_{\rm PH}$, and the ISCO radius $r_{\rm ISCO}$, respectively.

Overall, for the same parameter combinations, the minimum black hole spin required for energy extraction in the plunging region is significantly lower than that needed for the circular orbit region. This property implies that for observed low-spin black holes exhibiting energy extraction signatures, the process most likely originates from the plunging region, rather than the circular orbit region. On the other hand, in the plunging region, the inner spatial limit for energy extraction is no longer the photon sphere, but the event horizon. A comparison of the respective subplots in Figs.~\ref{CO_Region} and~\ref{PO_Region} shows that, the effect of all parameters follows the same pattern for both regions. One prominent difference is that when the plasma plunges within the ISCO, the energy extraction region gradually changes its shape as the orientation angle $\xi$ increases, rather than undergoing a pure reduction in size [see Figs.~\ref{CO_Region_different_xi} and~\ref{PO_Region_different_xi}]. Furthermore, as shown in Figs.~\ref{CO_Region_different_ell} and~\ref{PO_Region_different_ell}, variations in the KR field have a less pronounced impact on the energy extraction region in the plunging region compared to the circular orbit region. Lastly, since the spin range allowing energy extraction from the plunging region is broader than that in the circular orbit region, the constraints on the KR field derived from the plasma in the plunging region are generally less stringent than those obtained from the plasma in the circular orbit region.

\begin{figure*}[t]
	\centering
	\subfigure[]{\includegraphics[width=0.45\linewidth]{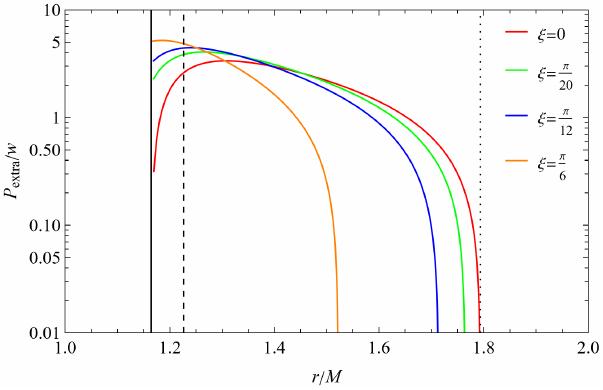}\label{PO_Power_different_xi}}
	\subfigure[]{\includegraphics[width=0.45\linewidth]{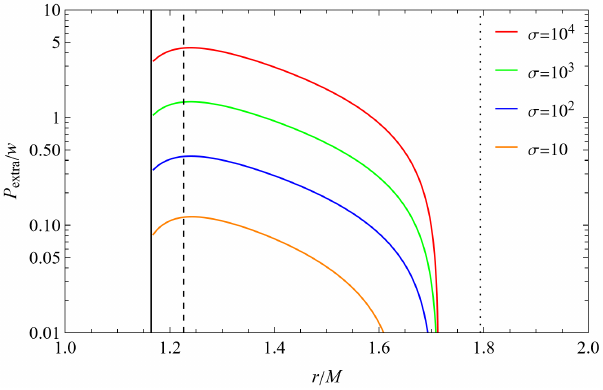}\label{PO_Power_different_sigma}}
	\\
	\subfigure[]{\includegraphics[width=0.45\linewidth]{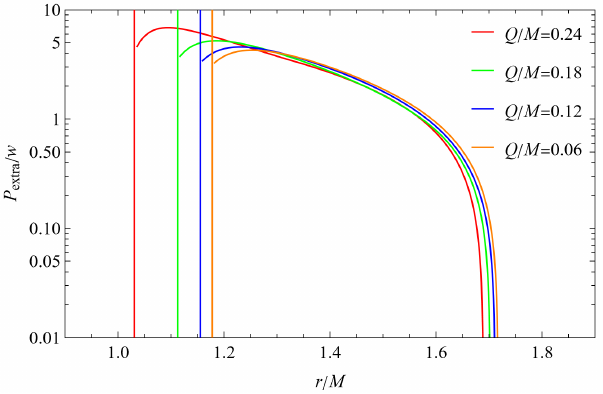}\label{PO_Power_different_Q}}
	\subfigure[]{\includegraphics[width=0.45\linewidth]{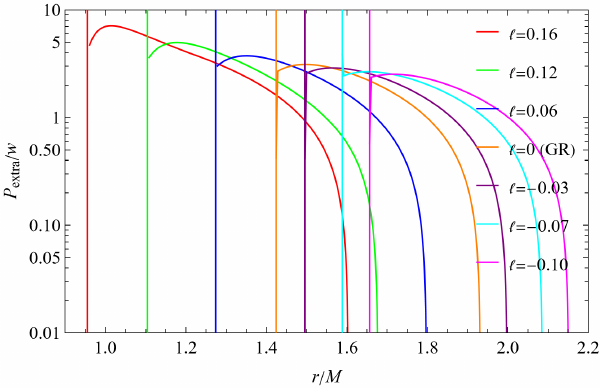}\label{PO_Power_different_ell}}
	\caption{Energy extraction power per unit enthalpy density $P_{\rm extra}/w$ in the plunging region for different parameter combinations. The vertical solid lines indicate the position of the event horizon. The parameter combinations are identical to those used in Fig.~\ref{CO_Power}.}
	\label{PO_Power}
\end{figure*}

Next, we discuss the energy extraction power per unit enthalpy density $P_{\rm extra}/w$ in the plunging region. The bulk plasma initially follows circular orbit outside the ISCO and subsequently plunges into the black hole, so the cross sectional area of the inflowing plasma depends on the ergosphere boundary and the event horizon:
\begin{equation}
    A_{\rm in} \sim r_{\rm E}^{2} - r_{\rm H}^{2} .
\end{equation}
To facilitate comparison with the circular orbit region, we use the same parameter combinations as in Fig.~\ref{CO_Power}. The power in the plunging region is still calculated using Eq.~\eqref{power}, with the results presented in Fig.~\ref{PO_Power}. When compared to Fig.~\ref{CO_Power}, the power in the plunging region shows notable differences, which can be summarized in three key aspects. First, for all parameters, the inner spatial boundary of the power is no longer the photon sphere, but the event horizon. Second, the initial power in the plunging region is non-zero at the event horizon. Third, the effects of the parameters $\xi$ and $Q$ on the power are more complex. Furthermore, a comparison of Figs.~\ref{CO_Power_different_ell} and~\ref{PO_Power_different_ell} shows that in the plunging region, the influence of the KR field on the power is less pronounced than that in the circular orbit region. In Fig.~\ref{PO_Power_different_ell}, the power curves corresponding to different values of the  Lorentz-violating parameter $\ell$ only avoid overlapping when the differences in $\ell$ are sufficiently large. As a result, it is difficult to effectively constrain the KR field using the energy extraction power of the plasma in the plunging region.

\begin{figure*}[t]
	\centering
	\subfigure[]{\includegraphics[width=0.45\linewidth]{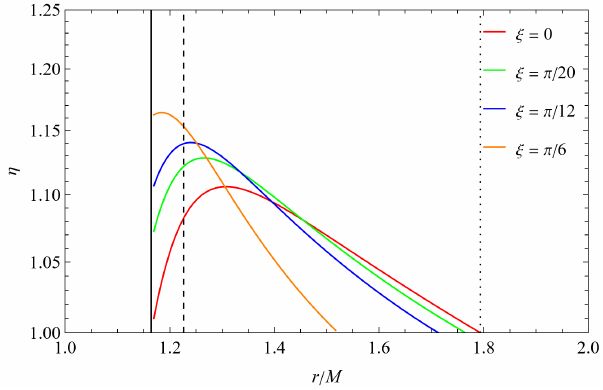}\label{PO_Eff_different_xi}}
	\subfigure[]{\includegraphics[width=0.45\linewidth]{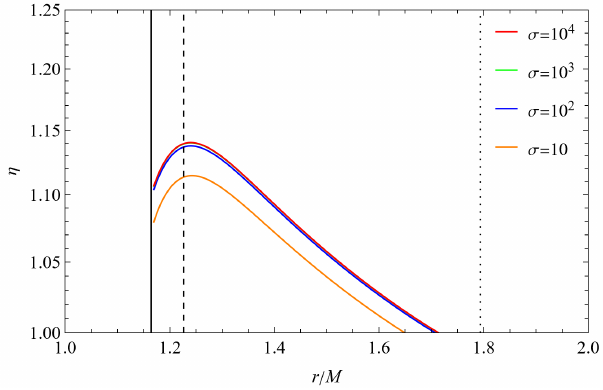}\label{PO_Eff_different_sigma}}
	\\
	\subfigure[]{\includegraphics[width=0.45\linewidth]{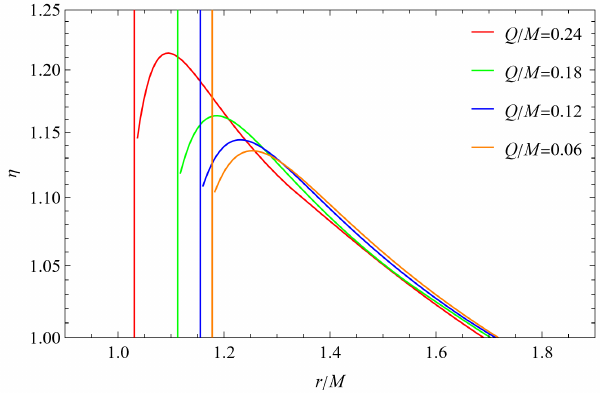}\label{PO_Eff_different_Q}}
	\subfigure[]{\includegraphics[width=0.45\linewidth]{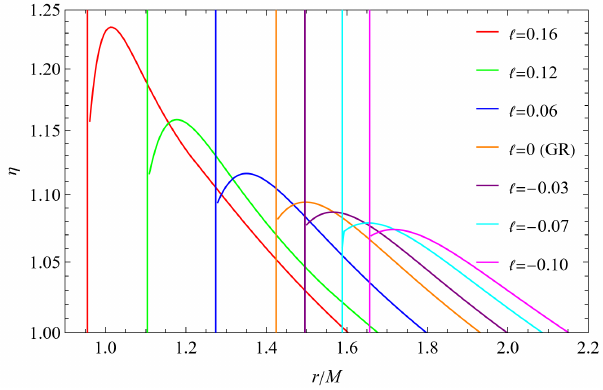}\label{PO_Eff_different_ell}}
	\caption{Energy extraction efficiency $\eta$ in the plunging region for different parameter combinations. The vertical solid lines indicate the position of the event horizon. The parameter combinations are identical to those used in Fig.~\ref{CO_Eff}.}
	\label{PO_Eff}
\end{figure*}

Finally, we present the energy extraction efficiency of the plasma in the plunging region in Fig.~\ref{PO_Eff}, using the same parameter combinations as in Fig.~\ref{CO_Eff}. A comparison of the two figures reveals that, for the two regions, the differences in the energy extraction efficiency closely resemble the differences in power, and therefore, they are not reiterated here. It is evident that constraining the KR field using the energy extraction efficiency of the plasma in the plunging region is less effective than that in the circular orbit region.

\section{Conclusion}\label{sec.5}

In this work, we have investigated energy extraction via the Comisso-Asenjo magnetic reconnection mechanism from rotating charged black holes in KR gravity. The KR field, as a fundamental field in string theory, could lead to low-energy effective Lorentz symmetry violation, a potential observational signature of quantum gravity. Our calculations demonstrate that such violation can leave computable imprints on the energy output characteristics of black holes. We focused on the impact of the KR field (i.e., the Lorentz-violating parameter $\ell$) on the characteristics of energy extraction, including the allowed parameter space, the energy extraction power and efficiency. A comparative analysis was conducted between the circular orbit region and the plunging region. Our principal conclusions can be summarized as follows.

Firstly, in the circular orbit region, the effects of the orientation angle $\xi$ and the plasma magnetization $\sigma$ on energy extraction are independent of the specific theory of gravity, and so the corresponding results are universally applicable~\cite{Comisso2021,Wei2022,Wang2022,Zhang:2024rvk,Li:2023htz,Zeng2025b,Li:2023nmy,Shen:2024sdr,Long:2024tws}. On the other hand, while the influence of the black hole charge $Q$ is significant, its trend shares commonalities across various charged black hole models~\cite{Zeng2025a,YuChih:2025hsg,Shaymatov:2023dtt}. We also found that the energy extraction process exhibits high sensitivity to the Lorentz-violating parameter $\ell$. This parameter not only significantly alters the allowed energy extraction region but also influences the energy extraction power and efficiency. Notably, as the parameter $\ell$ increases, the minimum black hole spin required for energy extraction decreases [see Fig.~\ref{CO_Region_different_ell}]. This characteristic suggests that observing the features of high-energy astrophysical jets or plasma outflows from black holes, such as their power and efficiency, could potentially provide observational constraints on the Lorentz-violating parameter $\ell$, thereby probing the KR field.

Secondly, in the plunging region, energy extraction presents a different picture compared to the circular orbit region. Here, particles plunge inward from the ISCO, acquiring a non-negligible radial velocity. This property extends the inner spatial boundary for energy extraction from the photon sphere all the way to the event horizon and drastically reduces the minimum black hole spin that allows energy extraction (see Fig.~\ref{PO_Region}). This finding carries important astrophysical implications: the high energy activity observed around low-spin black holes likely originates primarily from the plunging region. However, in the plunging region, the sensitivity of energy extraction to the Lorentz-violating parameter $\ell$ is markedly lower than that in the circular orbit region [see Figs.~\ref{CO_Region_different_ell} and~\ref{PO_Region_different_ell}]. Consequently, using observational data on the plasma in the plunging region to constrain the KR field offers weaker discriminative power.

Synthesizing the findings from both regions, we arrive at a central conclusion. The Lorentz-violating effects introduced by the KR field profoundly modify the dynamics and efficiency of black hole energy extraction via the Comisso-Asenjo magnetic reconnection mechanism. However, the potential for observational constraints critically depends on the region where magnetic reconnection occurs. For the circular orbit region, energy extraction provides a more sensitive ``probe'' for detecting the KR field or similar Lorentz-violating effects. In summary, this work reveals the rich characteristics of energy extraction from rotating charged black holes in KR gravity, which not only deepens our understanding of black hole energy release mechanisms within KR gravity but also provides a theoretical basis for future exploration of the KR field through observations of black holes.

\begin{acknowledgments}
This work is supported by the National Natural Science Foundation of China (Grants No.12505059, No.12505058, No.12505057, and No.12547101) and the China Postdoctoral Science Foundation (Grants No.2024M753825 and No.2025MD784184).
\end{acknowledgments}

\bibliographystyle{apsrev4-2}
\bibliography{reference}
\end{document}